\documentclass[aps,prb,twocolumn,showpacs,preprintnumbers,amsmath,amssymb,superscriptaddress,10pt]{revtex4-1}%
\usepackage{graphicx}
\usepackage{dcolumn}
\usepackage{amsmath}
\usepackage{color}
\usepackage{bm}

\newcommand{\tn}{\textit{T}$_{\rm N}$}
\newcommand{\tnfe}{\textit{T}$_{\rm N}^{\rm Fe}$}
\newcommand{\tnce}{\textit{T}$_{\rm N}^{\rm Ce}$}

\begin{document}
\title{Ferromagnetism and superconductivity in P-doped CeFeAsO}

\author{A. Jesche}
 \altaffiliation[]{present address: The Ames Laboratory, Iowa State University, Ames, USA}
 \email[]{jesche@ameslab.gov}
 \affiliation{Max Planck Institute for Chemical Physics of Solids, D-01187 Dresden, Germany}
\author{T. F\"orster}
  \affiliation{Max Planck Institute for Chemical Physics of Solids, D-01187 Dresden, Germany}
\author{J. Spehling}
  \affiliation{IFP, TU Dresden, D-01069 Dresden, Germany}
\author{M. Nicklas}
  \affiliation{Max Planck Institute for Chemical Physics of Solids, D-01187 Dresden, Germany}
\author{M. de Souza}
  \altaffiliation[]{present address: 
Departamento de F\'{i}sica, Universidade Estadual Paulista, Rio Claro (SP), Brazil}
  \affiliation{Physikalisches Institut, Goethe-Universit\"at Frankfurt, D-60438 Frankfurt(M), Germany}
\author{R. Gumeniuk}
  \affiliation{Max Planck Institute for Chemical Physics of Solids, D-01187 Dresden, Germany}
\author{H.~Luetkens}
  \affiliation{Laboratory for Muon Spin Spectroscopy, Paul Scherrer Institute, CH-5232 Villigen PSI, Switzerland}
\author{T. Goltz}
  \affiliation{IFP, TU Dresden, D-01069 Dresden, Germany}
\author{C. Krellner}
  \affiliation{Max Planck Institute for Chemical Physics of Solids, D-01187 Dresden, Germany}
\author{M. Lang}
  \affiliation{Physikalisches Institut, Goethe-Universit\"at Frankfurt, D-60438 Frankfurt(M), Germany}
\author{J. Sichelschmidt}
  \affiliation{Max Planck Institute for Chemical Physics of Solids, D-01187 Dresden, Germany}
\author{H.-H. Klauss}
 \affiliation{IFP, TU Dresden, D-01069 Dresden, Germany}
\author{C. Geibel}
  \affiliation{Max Planck Institute for Chemical Physics of Solids, D-01187 Dresden, Germany}

\begin{abstract}
We report on superconductivity in CeFeAs$_{1-x}$P$_{x}$O and the possible coexistence with Ce-ferromagnetism (FM) in a small homogeneity range around $x = 30\,\%$
with ordering temperatures of $T_{SC} \cong T_C \cong 4$\,K. 
The antiferromagnetic (AFM) ordering temperature of Fe at this critical concentration is suppressed to \textit{T}$_{\rm N}^{\rm Fe}\,\approx 40$\,K and does not shift to lower temperatures with further increase of the P concentration. 
Therefore, a quantum-critical-point scenario with \textit{T}$_{\rm N}^{\rm Fe}\,\rightarrow 0$\,K which is widely discussed for the iron based superconductors can be excluded for this alloy series. 
Surprisingly, thermal expansion and X-ray powder diffraction indicate the absence of an orthorhombic distortion despite clear evidence for short range AFM Fe-ordering from muon-spin-rotation measurements. 
Furthermore, we discovered the formation of a sharp electron spin resonance signal unambiguously connected with the emergence of FM ordering. 
\end{abstract}

\maketitle

One interesting aspect of the iron arsenides is the large variety of possible manipulations that result in superconductivity starting from antiferromagnetically (AFM) ordered parent compounds.
Taking LaFeAsO as an example, superconductivity (SC) can be induced by pressure \cite{Okada2008}, oxygen vacancies \cite{Ren2008} or substitution of any of the elements, e.g., La $\rightarrow$ Sr\,\cite{Wen2008}, Fe $\rightarrow$ Co\,\cite{Sefat2008}, As $\rightarrow$ P\,\cite{Wang2009}, or O $\rightarrow$ F\,\cite{Kamihara2008}.
The suppression of AFM ordering of Fe is leading to a superconducting ground state in all of these cases.
This is a very general property of the Fe-based superconductors and in fact there are only a few examples known where SC does not show up despite vanishing Fe ordering 
(e.g. Mn-doped LaFeAsO\,\cite{Berardan2009} or BaFe$_{2}$As$_{2}$\,\cite{Kim2010}).
In this sense CeFeAs$_{1-x}$P$_x$O presents a unique situation because 1$^{\rm st}$: replacing As by P 'usually' results in SC as shown for \textit{A}Fe$_{2}$As$_{2}$ (\textit{A} = Ca\,\cite{Shi2010}, Ba\,\cite{Jiang2009}, Sr\,\cite{Shi2010}, Eu\,\cite{Jeevan2010}) and \textit{R}FeAsO (\textit{R} = La\cite{Wang2009}, Sm\,\cite{Li2010}) and 2$^{\rm nd}$: CeFeAsO becomes superconducting when substituting Fe by Co\,\cite{Zhao2010} or O by F\,\cite{Chen2008} and also by inducing oxygen vacancies\,\cite{Ren2008}.
But combining P-doping with \textit{R} = Ce among the \textit{R}FeAsO compounds results in exceptional behavior: the Fe-ordering is suppressed as shown by de la Cruz \textit{et al.}\,\cite{delaCruz2009} and Luo\,\textit{et al.}\,\cite{Luo2010}, however, no superconductivity has been observed yet in CeFeAs$_{1-x}$P$_x$O.
Instead a crossover to ferromagnetic (FM) ordering of Ce was reported at the critical concentration of $x \approx 40\,\%$ where the AFM ordering of Fe vanishes\,\cite{Luo2010}.  
In contrast or in addition to this results we have found: 
(1) long range FM ordering coexisting with superconductivity in a small homogeneity range around $x = 30\,\%$, (2) evidence against a quantum critical scenario, (3) static AFM ordering of Fe moments in the absence of structural distortion, and (4) a clear electron spin resonance (ESR) signal connected to ferromagnetism.

Poly- and single crystalline material was synthesized as described in\,\cite{Jesche2010}. Phosphorous concentrations $x$ are given in nominal values. Energy dispersive X-ray analysis and X-ray powder diffraction (using Vegards rule) set an upper limit for the error in $x$ of 8\%. 
However, the continuous development of physical properties with $x$ implies that the deviation is significantly smaller.
Electrical resistivity of single crystals was measured along the $ab$-plane in 4-point geometry using a Quantum Design PPMS. 
Pressure experiments on single crystals were performed with a piston-cylinder cell using silicon oil as pressure-transmitting medium. 
Magnetization was measured in a Quantum Design MPMS. 
The thermal expansion coefficient, $\alpha(\textit{T})=\textit{l}^{-1}(\partial \textit{l}/\partial \textit{T})$, was measured on single crystals using a high-resolution capacitive dilatometer\,\cite{Pott1983}. 
X-ray powder diffraction (XRD) was performed at ESRF Grenoble (beamline ID 31, $\lambda = 0.39987$\AA).
Muon spin rotation and relaxation ($\mu$SR) experiments on polycrystalline samples were performed at S$\mu$S at the Paul Scherrer Institute, Switzerland (beamline $\pi$M3).
ESR was measured on polycrystalline material at 9.4 GHz using a standard spectrometer together with a He-flow cryostat.

\begin{figure}
\includegraphics[width=0.48\textwidth]{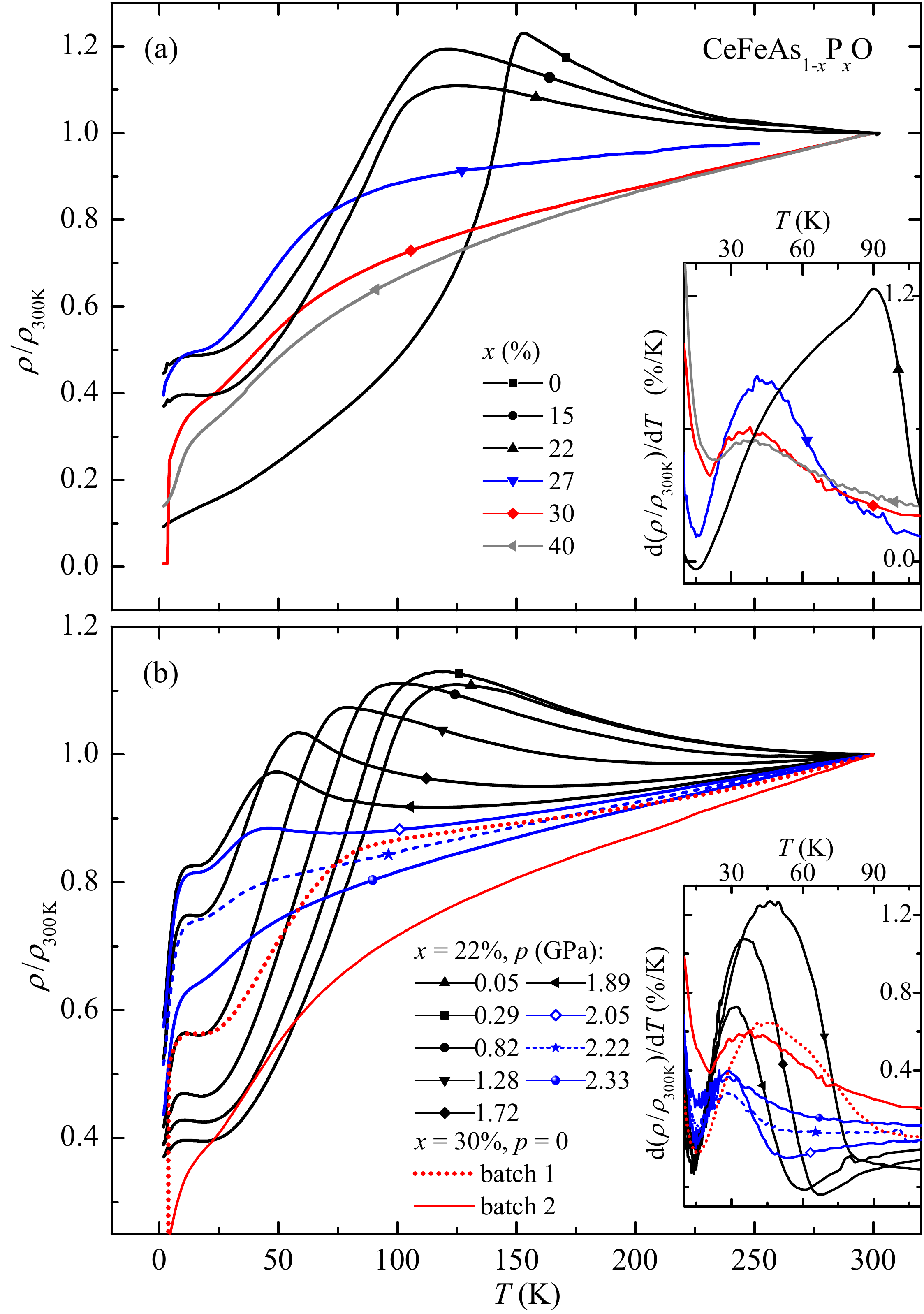}
\caption{(Color online) Electrical resistivity along the ab-plane, normalized to room-temperature (RT), of CeFeAs$_{1-x}$P$_{x}$O single crystals for $0 \leq x \leq 40\,\%$ at (a) ambient pressure and (b) for $x = 22\,\%$ as a function of pressure. Above a critical concentration of $x \approx 27\,\%$ and a critical pressure of $p \approx 2$\,GPa \tnfe~does not shift to lower temperatures anymore, instead the signature fades away. CeFeAs$_{0.70}$P$_{0.30}$O (red lines) shows clear signatures of the Fe-ordering at $T \approx 40$\,K and the onset of superconductivity at $T_{\rm SC} = 4$\,K. Insets: derivatives showing clear anomalies at \tnfe.}
\label{fig1}
\end{figure}

Fig.\,\ref{fig1} shows the suppression of the Fe-AFM ordering in P-doped CeFeAsO (a) and in CeFeAs$_{0.78}$P$_{0.22}$O under hydrostatic pressure (b). 
The P-doping, acting as chemical pressure, leads to an effective suppression of \tnfe~similar to Co- or F-doping. Whereas a broad superconducting dome was observed for the latter one, CeFeAs$_{1-x}$P$_{x}$O shows zero resistivity only in a narrow concentration range around $x \cong 30\,\%$, cf.~Fig.\,\ref{fig4}.
A clear signature of the Fe-ordering is visible in the derivative of the resistivity for all P-concentrations (inset in Fig.\,\ref{fig1}a, $x \leq 40\,\%$). 
For CeFeAs$_{0.78}$P$_{0.22}$O \tn~is already suppressed to $T = 100$\,K  at ambient pressure (compared to \tnfe~= 145\,K of undoped CeFeAsO, as determined by the maximum in the first derivative of the electrical resistivity). Tracing this signature allows to study the evolution of \tnfe~at low temperatures using rather small pressures. 
Increasing the pressure first shifts \tnfe~to lower temperatures until above $p \approx 2$\,GPa the signature becomes less pronounced. However, \tnfe~remains constant as seen from the 
constant peak position in the derivative (inset in Fig.\,\ref{fig1}b).
The normalized resistivity at $T = 15$\,K first increases with increasing pressure followed by a decrease for $p > 2$\,GPa. 
This corresponds to a weakening of the AFM-Fe ordering
for increasing pressure, because the
increasing fluctuations of Fe-moments result
in higher resistivities in the vicinity of the transition from a magnetically ordered to a paramagnetic Fe ground state. 

Samples with a higher P-concentration of $x = 30\,\%$ show the onset of superconductivity at $T_{SC} = 4$\,K together with signatures of the Fe-AFM ordering at \tnfe $\,\approx 40$\,K (shown in Fig.\,\ref{fig1}b for two different samples). 
A further increase of $x$ to 40\,\% does not shift \tnfe~below 35\,K. Instead the signature fades away, similar to $x = 22\,\%$ under hydrostatic pressure (see phase diagram in Fig.\,\ref{fig4}d).
This behavior under both chemical and hydrostatic pressure renders a quantum critical scenario with a continuous suppression towards \tnfe $\,\rightarrow 0$\,K unlikely.

\begin{figure}
\includegraphics[width=0.483\textwidth]{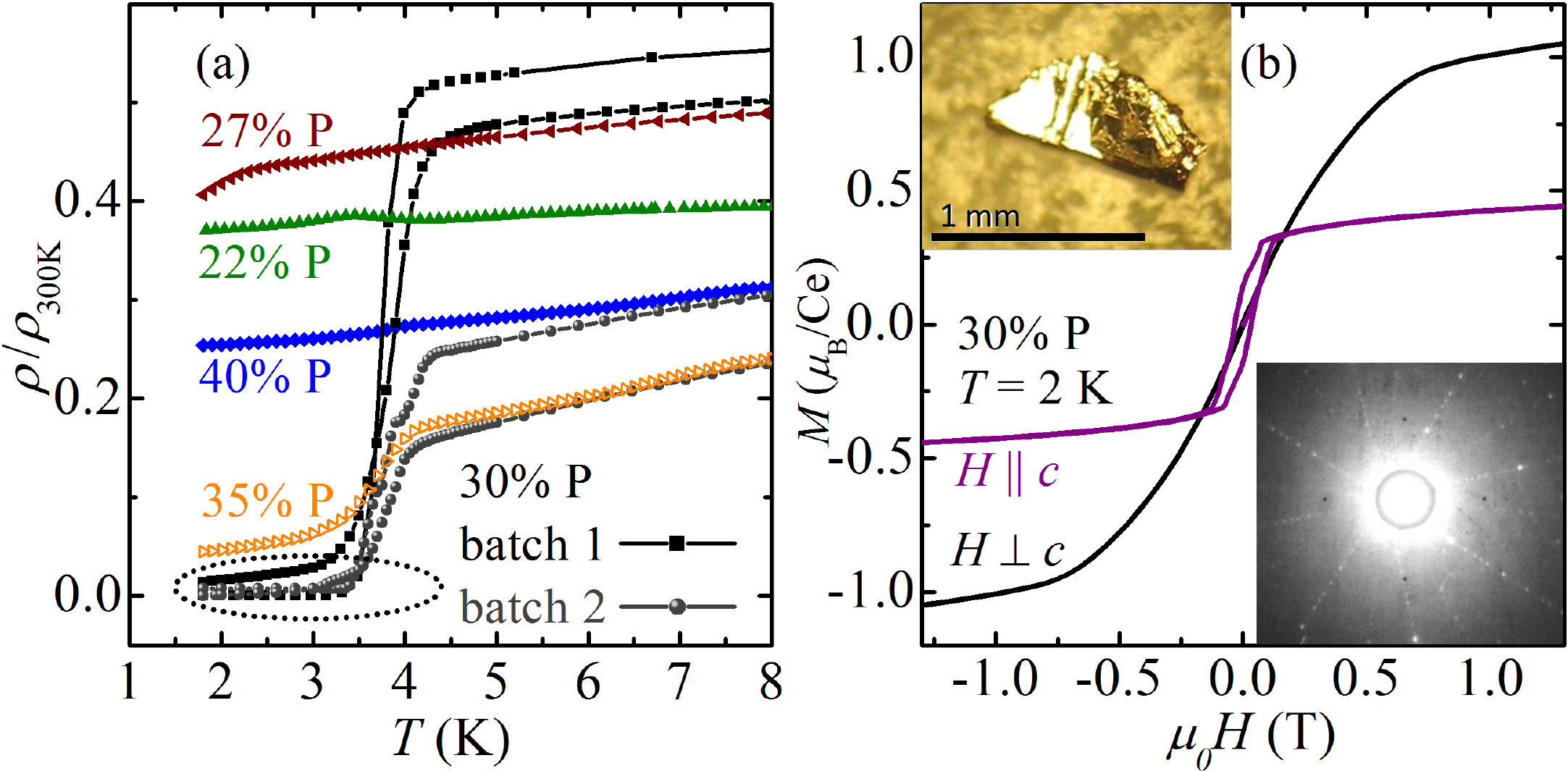}
\caption{(Color online) a) Electrical resistivity of single crystalline CeFeAs$_{1-x}$P$_{x}$O normalized to RT. Zero resistivity is observed for $x = 30\,\%$ (plotted for four different samples of two different batches) whereas other samples show only weak ($x = 27\,\%, 35\,\%$) or absolutely no indications ($x = 22\,\%, 40\,\%$) for SC. b) Anisotropic magnetization of CeFeAs$_{0.70}$P$_{0.30}$O showing FM behavior with the spontaneous moment along the $c$-axis. Note: the same single crystal shows zero resistivity (lower curve of batch 2 in the left panel). The single crystal used for the measurements (upper left) and a corresponding Laue back reflection pattern (lower right) are displayed.}
\label{fig2}
\end{figure}

Electrical resistivity at low temperatures is shown in Fig.\,\ref{fig2}a for several P-concentrations. 
Zero resistivity is observed only for $x = 30\,\%$ (in 5 of 6 samples), in contrast to all samples with $x \neq 30\,\%$ (with one exception for 1 of 3 samples with $x = 35\,\%$). 
To check for reproducibility, the crystal growth for $x = 30\,\%$ has been repeated (indicated as batch 2 in Fig.\,\ref{fig2}) and revealed a similar result.
From the distinct behavior of samples with slightly different $x$ values the inhomogeneity of the P-concentration can be inferred to be small - note the absence of SC for $x = 27\,\%$!

The emergence of SC in CeFeAs$_{1-x}$P$_{x}$O is correlated with a weakening of the AFM-Fe ordering at $x = 30\,\%$. At this critical concentration the magnetic ordering of Ce suddenly changes from AFM to FM. We observed a well defined hysteresis with the spontaneous moment along the crystallographic $c$-axis clearly showing the FM ground state of this single crystalline sample (Fig.\,\ref{fig2}b). 
The size of the ordered moment is 0.33\,$\mu_B$/Ce. At $\mu_0H = 1$\,T the magnetization reaches values of $\mu_{\rm sat}^c = 0.43\,\mu_B$/Ce for $H \parallel c$ and $\mu_{\rm sat}^{ab} = 1.00\,\mu_B$/Ce for $H \perp c$ with $\mu_{\rm sat}^c$ increasing only slightly at higher fields. 
This indicates an ordering of local Ce$^{3+}$ moments in a $\Gamma_6$ crystal electric field ground state in the full volume of the sample with theoretical values for the saturation moment of 0.43\,$\mu_B$/Ce (along the $c$-axis) and 1.29\,$\mu_B$/Ce (along the $ab$-plane) as proposed for undoped CeFeAsO\,\cite{Jesche2009} and the isostructural compound CeRuPO\,\cite{Krellner2008}. 
The size of a FM component of the ordered moment for $x = 27\,\%$ and $x = 22\,\%$ (not shown) is smaller than 0.02\,$\mu_B$/Ce and 0.002\,$\mu_B$/Ce, respectively, again stressing the well defined P-concentration.

The simultaneous occurrence of FM and SC at $T_{\rm C} \cong T_{\rm SC} \cong 4$\,K precludes the proof of bulk superconductivity
because specific heat and magnetic susceptibility at this temperatures are completely dominated by the FM ordering of the Ce moments. 
A comparison with UCoGe, one of the few established FM superconductors, supports this possibility since the diamagnetic signal at $T_{\rm SC}$ can be small compared to the anomaly at $T_{\rm C}$\,\cite{Aoki2012} and a Meissner-phase might be even completely absent when the internal FM field is larger than $H_{c1}$\,\cite{Deguchi2010}. 
From the stabilization of FM ordering for $x > 30\,\%$ and the absence of SC in these samples we conclude a competing character of both effects (see phase diagram in Fig.\,\ref{fig4} and Luo\,\textit{et al.}\,\cite{Luo2010}).
However, the situation is more subtle than a simple competition of local moment ferromagnetism and conduction band superconductivity since the observed behavior of CeFeAs$_{0.70}$P$_{0.30}$O is significantly different from most superconducting iron pnictides already well above the Curie temperature of Ce, 
as we will show in the next section.

\begin{figure}
\includegraphics[width=0.48\textwidth]{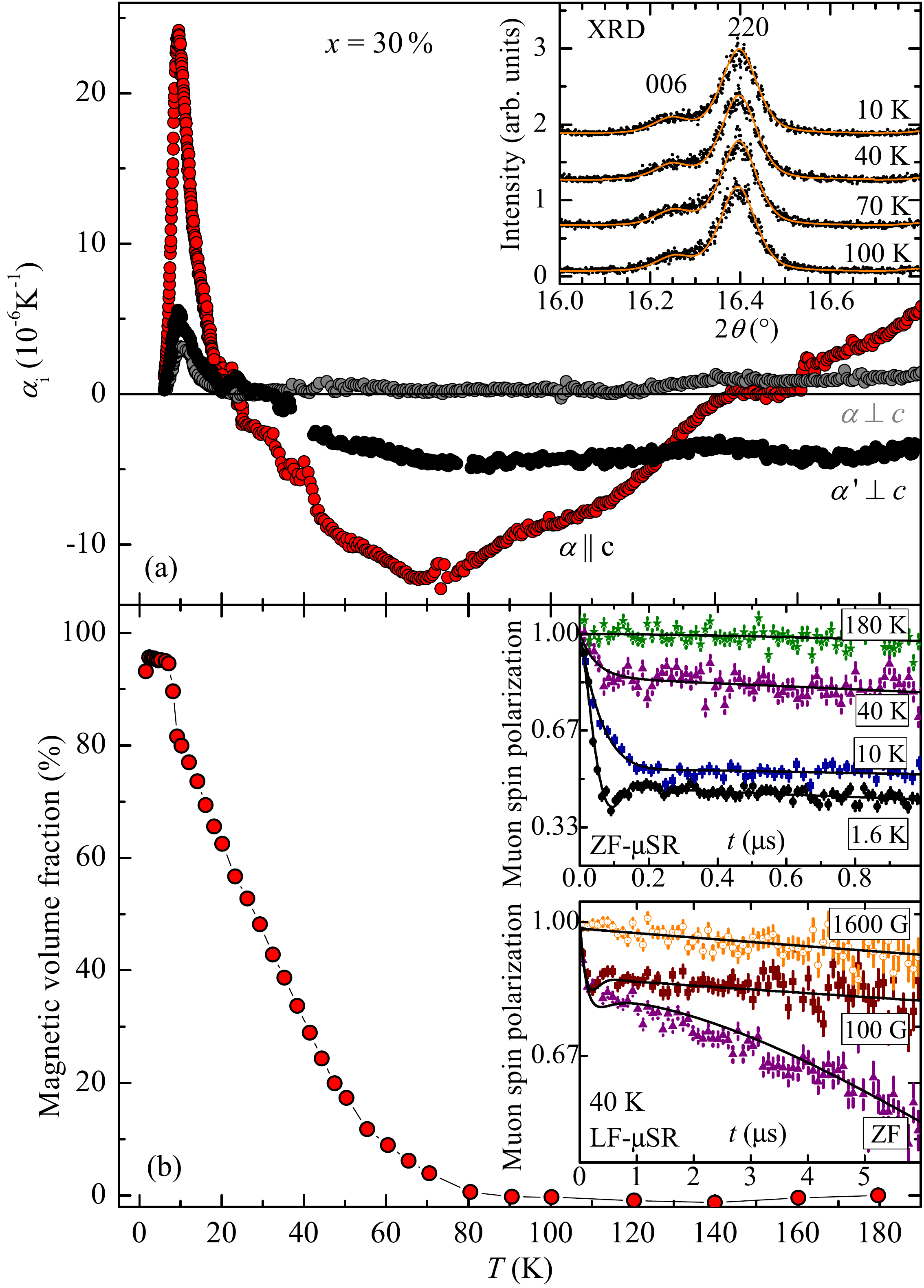}
\caption{(Color online) a) Anisotropic thermal expansion coefficients of CeFeAs$_{0.70}$P$_{0.30}$O showing no indication for an orthorhombic distortion \lbrack results for two different directions in the $ab$-plane, $\sphericalangle(\alpha,\alpha') \approx 60^{\circ}$\rbrack~ and a pronounced negative thermal expansion along the $c$-axis. 
The sharp peaks below 15 K are connected to the magnetic ordering of Ce.
Inset: synchrotron XRD pattern close to the 220 reflection - no indications for splitting or broadening are observable. b) Magnetic volume fraction determined from $\mu$SR measurements. Static magnetic ordering sets in below $T = 70$\,K. Upper inset: the strong relaxation of the muon spin polarization evidences short range magnetic ordering. Lower inset: longitudinal field $\mu$SR measurements proving the predominantly static character of the magnetism.}
\label{fig3}
\end{figure}

Fig.\,\ref{fig3}a shows the thermal expansion coefficient, $\alpha_i$, measured along and perpendicular to the $c$-axis. In contrast to undoped CeFeAsO \cite{Jesche2010}, $\alpha_{\perp c}$ is small and almost constant for 20\,K $ < T < 190$\,K.
In particular it lacks any indication for a structural transition towards the orthorhombic phase. Note that thermal expansion is very sensitive to structural phase transitions, usually more sensitive than XRD, due to its higher resolution. These results are consistent with the XRD pattern which show neither a splitting nor a broadening of the
220 reflection (inset in Fig.\,\ref{fig3}a). 
Along the $c$-axis a broad negative $\alpha_{\parallel c}$ anomaly, centered around 70\,K, is observed which is related to the onset of Fe ordering as evidenced by a comparison with the $\mu$SR results (see below).
Between $T = 10$\,K and 100\,K the length change amounts to $\Delta l/l_{\parallel c} = -6.9\cdot10^{-4}$ which is in good agreement with the value of $\Delta l/l_{\parallel c} = -5.3\cdot10^{-4}$ determined from XRD on polycrystalline material. 

Fig.\,\ref{fig3}b shows the developement of the magnetic volume fraction of CeFeAs$_{0.70}$P$_{0.30}$O as determined from weak transverse field $\mu SR$ measurements. 
We find that static magnetic order gradually develops below $T = 70$\,K down to 10 K, as displayed in the main panel. 
In the upper inset of Fig.\,\ref{fig3}b we show representative zero-field (ZF) $\mu$SR data at 1.6, 10, 40, and 180 K. 
At high $T$, the muon spin polarization $P(t)$ is well described by the Gaussian Kubo-Toyabe (GKT) function $P(t)=\frac{1}{3}+\frac{2}{3}[1-(\sigma t)^{2}]\exp(\frac{1}{2}\sigma^{2}t^{2})$, as expected for static and randomly oriented magnetic fields originating from nuclear moments only~\cite{Hayano1979}. 
This means that at 180\,K CeFeAs$_{0.70}$P$_{0.30}$O is in the paramagnetic (PM) state and that the electronic Fe-3$d$ and Ce-4$f$ moments are rapidly fluctuating so that the resulting muon depolarization is small compared with the nuclear contribution.
Below $T\approx70$ K, an additional exponential depolarization gradually develops with decreasing $T$ on the cost of the PM GKT signal. 
The exponential behavior indicates that this relaxation stems from static or slowly fluctuating electronic moments. 
Longitudinal field (LF) $\mu$SR experiments can distinguish between these two scenarios\,\cite{Hayano1979}. 
Our LF-$\mu$SR measurements at 40\,K prove the predominantly static character of the internal magnetic fields most likely due to short-ranged ordered Fe-3$d$ electronic moments with a small dynamic contribution 
as displayed in the lower inset of Fig.\,\ref{fig3}b. 

Therefore, the data shown in Fig.\,\ref{fig3} give evidence for static magnetic ordering of Fe in the absence of an orthorhombic distortion - a scenario which is in contrast to the widely discussed nematic model for Fe-based superconductors. 
On the other hand: the development of static AFM order in CeFeAs$_{1-x}$P$_{x}$O instead of AFM fluctuations existing in a nematic phase might be the origin for the absence of high-temperature SC in this alloy series. 

\begin{figure}
\includegraphics[width=0.48\textwidth]{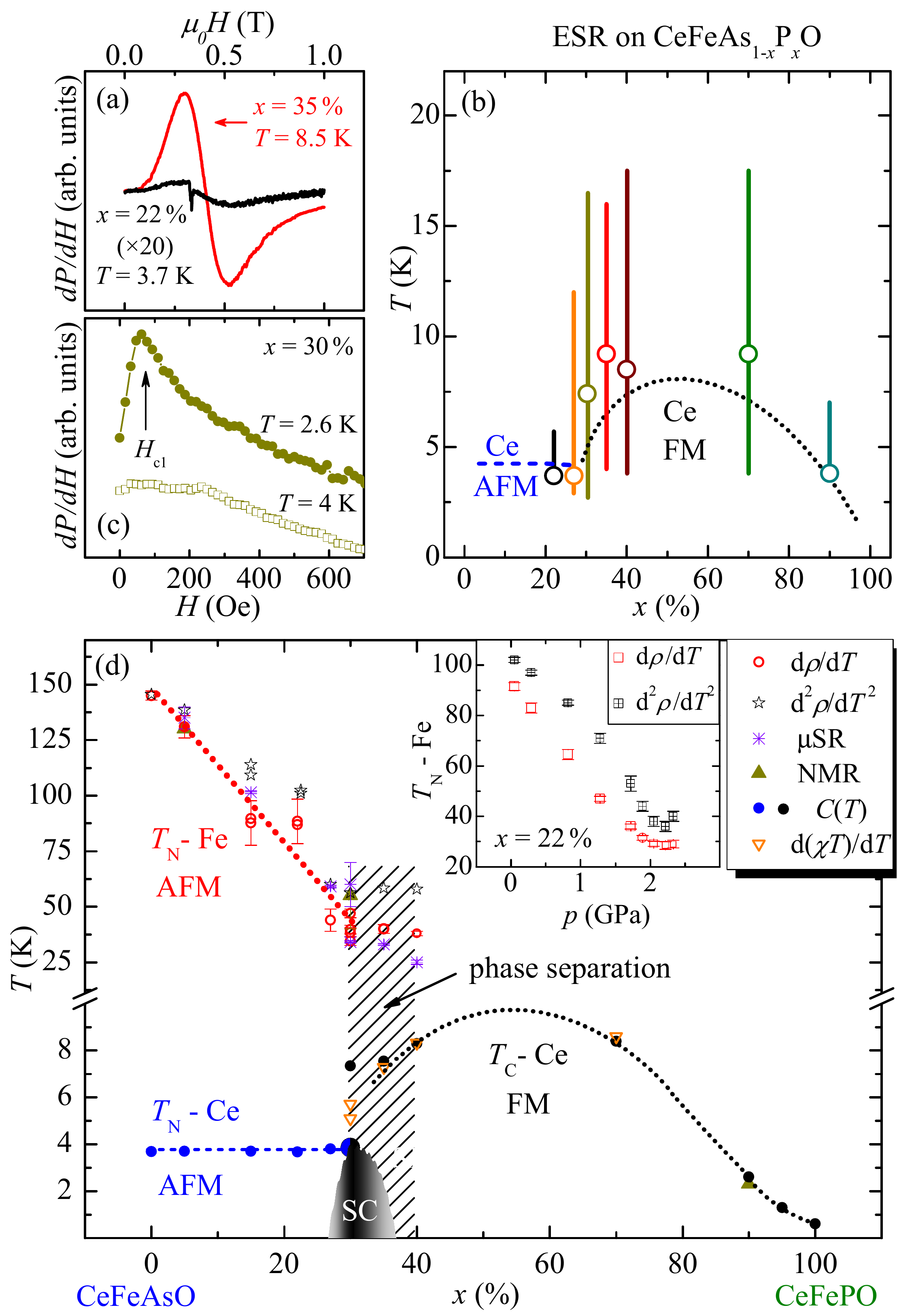}
\caption{(Color online) a) Sharp ESR signal for $x = 35\,\%$ compared to a weak signal for $x = 22\,\%$ (multiplied by a factor of 20 for visibility). b) Observability range of the ESR signal as a function of $x$ indicated by vertical lines with the strongest signals indicated by circles. No ESR line was observed for $x < 22\,\%$ and $x > 90\,\%$. c) Strong increase of the absorption at small fields indicates SC with a lower critical field of $H_{c1} = 70$\,Oe which was solely observed for $x = 30\,\%$ and $T < 4$\,K. d) Characteristic temperatures as a function of $x$ determined by several techniques. The inset shows \tnfe~for $x = 22\,\%$ as a function of pressure.}
\label{fig4}
\end{figure}

Finally we report the discovery of a sharp ESR signal for a certain range of $x$ which confirms the relevance of FM fluctuations for its observation \cite{Krellner2008c}. 
Figure\,\ref{fig4}a shows the ESR signal for $x = 35\,\%$ at $T = 8.5$\,K (FM ground state) compared to the weak signal for $x = 22\,\%$ (AFM ground state, measured on samples with similar mass and shape).
The $T$-$x$-range where an ESR signal was observed is shown in Fig.\,\ref{fig4}b. 
There is an obvious relation to FM fluctuations occurring at $T > T_C$ in samples with a FM ground state (see phase diagram in Fig.\,\ref{fig4}d).
No ESR signal was observed for $x < 22\,\%$ and $x > 90\,\%$ in measurements on several poly and single crystals.
The characteristics of this ESR signal are similar to those reported for the ferromagnet CeRuPO \cite{Foerster2010}, suggesting that this ESR signal stems from correlated $4f$-electrons.
Furthermore, ESR measurements provide further evidence for SC in CeFeAs$_{0.70}$P$_{0.30}$O. Only in these samples the absorption at low field increases strongly for $T < T_{\rm SC}$ 
due to non-resonant absorption caused by flux flow
(Fig.\,\ref{fig4}c). A lower critical field of $H_{c1} = 70$\,Oe was estimated from the maximum in d$P$/d$H$.

The main results are summarized in the phase diagram in Fig.\,\ref{fig4}d. Transition temperatures are determined by means of electrical resistivity (1st and 2nd derivative), $\mu$SR (50\% magnetic volume fraction), NMR\,\cite{Sarkar2012} (line broadening), specific heat, and magnetic susceptibility \lbrack minimum in d$(\chi T)$/d$T$\rbrack. Whereas \tnfe~decreases monotonically with increasing P-concentration \tnce~is almost constant for $x < 30\,\%$. 
The vanishing of \tnfe~at a finite temperature together with the crossover from AFM to FM ordering of Ce and the emergence of SC give rise to a possible tricritical point at finite temperature at $x = 30\,\%$, a scenario which has been theoretically discussed for LaFeAsO$_{1-x}$F$_x$ by Giovannetti\,\textit{et al.}\,\cite{Giovannetti2011}.
Signatures of the Fe-ordering are still observable for $x > 30\,\%$, however the transition temperature seems to stay above a certain threshold 
 and the anomalies in $\rho(T)$ are getting less pronounced suggesting phase separation associated with a first order transition. Similar behavior is observed for CeFeAs$_{0.78}$P$_{0.22}$O under external pressure (inset Fig.\,\ref{fig4}d) where the maximum pressure applied might be too low to induce SC. 

The suppression of the Fe-AFM ordering alone is not sufficient for the emergence of Ce-FM ordering as shown by F and Co-doping studies \cite{Zhao2010, Chen2008}. Therefore, the origin for this behavior is more complex than a possible simple sign change of the RKKY interaction and will hopefully stimulate further theoretical investigations. 
For $x > 70\,\%$ the FM ordering of Ce becomes weaker and for $x = 90\,\%$, $T_C$ is reduced to 2.7 K.
P-concentrations of $90\,\% < x \leq 100\,\%$ result in further suppression of $T_C$ which will be in the focus of a forthcoming publication.

In summary, CeFeAs$_{1-x}$P$_x$O was found to be one of the rare examples showing a close proximity of superconductivity and ferromagnetism. 
The formation of static Fe-ordering in the absence of an orthorhombic distortion is the main difference to Fe-based high-T$_c$ superconductors and indirectly supports the relevance of nematic fluctuations for the emergence of higher critical temperatures. 

\section{Acknowledgments}
The authors would like to thank U. Schwarz, M. Sch\"oneich and C. Curfs for their support and help with the XRD experiment at ESRF, Grenoble and E. Lengyel for her help in setting up the pressure experiment.
G. Auffermann, U. Burkhardt and P. Scheppan are acknowledged for chemical analysis of the samples. 
This work was supported by the SPP 1458 priority program of the Deutsche Forschungsgemeinschaft.

\end{document}